\documentclass[11pt]{article}
\usepackage{pdflscape}
\usepackage{graphicx}
\usepackage{multirow}
\usepackage{dcolumn}
\usepackage{natbib}
\usepackage{amsmath}
\usepackage{amssymb}
\usepackage[T1]{fontenc}
\usepackage[utf8]{inputenc}
\usepackage{xcolor}
\usepackage{float}
\usepackage{subcaption}
\usepackage{booktabs}
\usepackage{longtable}
\usepackage{threeparttable}
\usepackage{hyperref}
\hypersetup{colorlinks,linkcolor={blue},citecolor={blue!50!black},urlcolor={blue}}
\usepackage{geometry}
\usepackage{authblk}
\usepackage{setspace}
\begin{document}
\title{\textbf{Regional advantage in rugby sevens:} \\ \textbf{Is there a home effect when nobody is home?}
\thanks{Federico Fioravanti acknowledges the financial support from the French National Research Agency within the project ANR-24-EXMA-0001 PEPR MathsVivES CONDORCET.}
}
\author[1]{Fernando Delbianco}
\author[2]{Federico Fioravanti\footnote{federico.fioravanti@univ-st-etienne.fr (corresponding author)}}
\author[1]{Fernando Tohm\'e}
\author[3]{Mart\'in Trombetta}
\affil[1]{Departamento de Economía, Universidad Nacional del Sur (UNS)-Instituto de Matemática (INMABB)-CONICET, Bahía Blanca, Argentina.}
\affil[2]{Universit\'e Jean Monnet Saint-\'Etienne, CNRS, Universit\'e Lyon 2, emlyon business school, GATE, 42023, Saint-\'Etienne, France.}
\affil[3]{Instituto Interdisciplinario de Econom\'ia Pol\'itica (IIEP), Universidad de Buenos Aires--CONICET, Buenos Aires, Argentina}
\date{\vspace{-1pt}}

\maketitle

\begin{abstract}
\noindent We study the existence of a \emph{Regional Differential} in rugby sevens: whether, in tournaments where no competing team enjoys formal home status, some national sides systematically over- or under-perform depending on \emph{where} the event is staged. Using the universe of $2{,}672$ men's and women's matches from international rugby sevens tournaments played between 2016 and 2025, principally the World Rugby Sevens Series, but also the World Cup Sevens and the Olympic Games, we replace the binary ``locality'' indicator of the classic home advantage literature with continuous measures of geographic, temporal, and cultural proximity between each team and the host country. We show that aggregate associations between proximity and performance (which appear large and highly significant in naive specifications) are almost entirely an artifact of \emph{team composition} and \emph{scheduling}: once team and opponent quality are absorbed through fixed effects in a symmetric team-match panel, no general regional advantage survives. However, this null aggregate masks substantial heterogeneity. For a well-defined subset of teams (most notably France, but also New Zealand, the United States, Canada, Spain, Ireland, Kenya, Samoa, and Brazil), performance declines significantly and monotonically with the distance separating the host city from home. Some of the established southern-hemisphere powers (Fiji, South Africa, Australia, Argentina) are essentially distance-neutral. We further show that for some teams the gradient operates through east-west jet lag rather than pure displacement. The Regional Differential in sevens is therefore real but \emph{team-specific} rather than universal, a distinction obscured by pooled estimation.

\vspace{4pt}
\end{abstract}

\section{Introduction}

The Home Advantage (HA) effect in sports, defined as the tendency for the home team to achieve superior results, has been widely documented across numerous disciplines \citep{jones2007home, thomas2008home, nevill1997identifying, jamieson2010home}. Although its underlying mechanisms remain debated, prior research points to several potential drivers, including physiological responses \citep{neave2003testosterone}, psychological influences \citep{agnew1994crowd}, economic factors \citep{carmichael2005home}, and officiating biases \citep{downward2007effects}. Among these, the presence of a supportive home crowd emerges as particularly influential.

In football, for instance, multiple studies report a significant decline in HA when matches are played behind closed doors \citep{bryson2021causal,bilalic2021home,mccarrick2021home,leitner2022cauldron}, with similar findings in basketball, American football, and rugby \citep{leota2022home,szabo2022impact,delbianco2023home}. Recent evidence from Argentina's football league further suggests that the presence of local supporters matters more than that of away fans \citep{fioravantivisitorsout2025}.

Much less is known, however, about HA in contexts where \emph{no} team enjoys home status. Such situations arise in international competitions like the FIFA World Cup, where matches are staged at neutral venues, often without a host-country representative. A particularly interesting case, and the focus of our study, is international rugby sevens, encompassing the World Rugby Sevens Series,\footnote{Although the competition's name has varied over time, we use ``World Rugby Sevens Series'' as its most representative and current designation.} the World Cup Sevens, and the Olympic Games.

The bulk of our dataset consists of World Rugby Sevens Series matches. This annual series, organized by World Rugby, features national teams competing across multiple international tournaments. While the format has evolved over time, its core design has remained stable: tournaments are held worldwide, typically running from November to June, with teams accumulating points based on their finishing position at each event. Although the team with the highest accumulated points at season's end is declared the overall champion, each individual tournament carries its own trophy, giving teams a standing incentive to win every leg of the series, not merely to accumulate points. Since 2023, the men's and women's competitions have been staged at shared venues. Crucially, most matches do not involve a host-country team, and some hosts (such as Singapore and the United Arab Emirates) have never fielded a national side at all.

The other two competitions in our sample, the World Cup Sevens and the Olympic Games, are single-event tournaments with a format similar to that of individual legs of the World Rugby Sevens Series. We therefore treat teams' incentives to win as comparable across all three competitions.

Tournament scheduling is designed to contain economic costs, but it also has the incidental effect of neutralizing travel-based advantages. Events are typically staged over a single weekend, with teams arriving several days earlier so as to shed jet lag and acclimatize to local conditions. Consecutive rounds are, moreover, often geographically clustered, so that all teams face broadly comparable itineraries. Beyond its economic efficiency, this arrangement should in principle limit the edge that any single team might otherwise gain from a shorter journey. Our results suggest that it largely succeeds in the aggregate, but falls short for an identifiable subset of teams.

Despite the absence of formal ``home'' teams, it is widely believed that certain sides perform systematically better at particular venues. Fiji and New Zealand, for example, are reputed to excel in Hong Kong or Japan, while Argentina is thought to travel well to the United States, Canada, or Australia. Our first objective is therefore to assess whether specific teams systematically over- or under-perform in particular host countries. We refer to this pattern as the \emph{Regional Differential}: the influence of the host region on team performance, net of team strength. Where such effects are present, we explore possible explanations.

Traditional HA studies model ``locality'' as a binary variable, equal to $0$ or $1$. In tournaments without a host team, however, this framework is overly restrictive, if not meaningless. An alternative is to treat locality as a continuum determined by the degree of proximity between a team and the host country. We consider two families of determinants, \emph{geographical} and \emph{cultural}. The first comprises raw distance and the east-west time zone displacement that produces jet lag, a channel for which there is direct evidence in sport: eastward travel and time zone shifts depress performance in baseball \citep{recht1995baseball}, basketball \citep{leota2022eastward}, and, in the NBA more broadly \citep{nutting2017timezones}, complementing the venue familiarity evidence of \citet{pollard2002stadium}.

The second family builds on a now-standard tradition in economics that treats cultural similarity between societies as something measurable. \citet{alesina2003fractionalization} construct ethnic, linguistic, and religious \emph{fractionalization} indices for roughly $190$ countries, each defined as the probability that two individuals drawn at random from a country belong to different groups along the corresponding dimension; these indices have since become a workhorse for relating heterogeneity to economic and institutional outcomes. A complementary literature moves from heterogeneity \emph{within} a country to \emph{distance between} countries, arguing that linguistic, religious, and broader cultural gaps act as frictions that impede the flow of goods, ideas, and trust across societies \citep{spolaore2013roots,spolaore2016ancestry}. We import both ideas into the neutral venue setting, using coarse binary indicators for whether a team shares the host's dominant language or religion, and a finer continuous measure given by the absolute difference between the team's country and the host in each \citet{alesina2003fractionalization} index, which compares the degrees of diversity of the two countries. If cultural proximity carries any of the advantage usually attributed to playing at home---a more sympathetic or familiar crowd, easier acclimatization, a larger diaspora in the stands---then teams should perform better at hosts that resemble them culturally, and their results should deteriorate as the cultural distance to the host widens. This lets us ask whether the cultural channel that the home advantage literature emphasizes survives once no team is formally ``home.''

Our central methodological point is that detecting such an effect requires careful attention to confounding factors. Because the series fields teams of vastly different quality, and because the geographic origin of teams is correlated with the venues they reach and the rivals they face, simple correlations between proximity and results are badly contaminated. We address this through a \emph{symmetric team-match panel} in which every match contributes one observation per participating team and the relevant proximity variables are defined from each team's own perspective. Team and opponent fixed effects then absorb the quality of both sides, while year and round fixed effects absorb calendar and tournament-stage effects.

Our findings can be summarized in three points. First, the \emph{aggregate} Regional Differential is essentially zero: the large, highly significant proximity coefficients found in specifications without quality controls collapse to insignificance once team and opponent fixed effects are introduced. Second, this aggregate null conceals economically meaningful heterogeneity: a clearly identified group of teams (led by France) performs worse the farther a tournament lies from home, while the traditional southern-hemisphere powers are distance-neutral (with the exception of New Zealand). Third, the mechanism is not uniform: for some teams the gradient reflects pure displacement, whereas for others (e.g.\ Canada) it operates through east-west time zone shifts. These results refine, rather than overturn, the folk wisdom about regional strongholds: such strongholds exist, but they are a property of \emph{particular} teams and are invisible in pooled estimation.

\section{Data and Methodology}\label{sec:datameth}

\subsection{Data}\label{sec:data}

Match-level data were retrieved from \url{https://sevensrugby.fr/} and cover international rugby sevens tournaments between 2016 and 2025. The raw sample comprises $2{,}672$ matches ($1{,}600$ men's and $1{,}072$ women's), played across $13$ host countries on six continents. For each match we observe the date, the two competing national teams, the final score, the host city and country, the tournament identifier, and the round (qualifying round, secondary table, main table).

Geographic distances between each team's reference city and the host city were computed with the \texttt{geopy} library on coordinates from the Natural Earth dataset.\footnote{\url{https://www.naturalearthdata.com/}} We additionally observe each location's UTC offset (used to construct jet lag), and the dominant language and religion of each team's country and of the host, from which we build binary cultural proximity indicators. For a finer cultural measure we also match each country to its ethnic, linguistic, and religious fractionalization indices \citep{alesina2003fractionalization}, which give the probability that two randomly drawn individuals in a country belong to different groups along each dimension.

Two features of the raw data shape our research design. First, the data list a ``Team~1'' and a ``Team~2'' per match, and Team~1 wins $67.3\%$ of the time: the listing is \emph{not} symmetric and is correlated with team strength. Naive regressions of the score difference on team-1 covariates therefore confound the variable of interest with which side happens to be listed first. Second, the geographic origin of a team is mechanically related to the set of venues it reaches and the opponents it meets there (the leading sides of Oceania, for instance, are simultaneously distant from northern venues \emph{and} unusually strong). Both features motivate the panel construction below.

\subsection{The symmetric team-match panel}\label{sec:panel}

We reshape the data into a team-match panel in which every match yields two observations, one for each participating team taken as the \emph{focal} side. For focal team $i$ facing opponent $j$ in match $m$ hosted in country $h$, the outcome variables are the score margin $y^{\text{margin}}_{im}=S_{i}-S_{j}$, a win indicator $y^{\text{win}}_{im}=\mathbf{1}[S_i>S_j]$, and a win-or-tie indicator. By construction the panel is symmetric: the mean margin is exactly $0$ and the win rate is $0.495$, eliminating the listing bias described above.

The proximity regressors are defined from the focal team's viewpoint: $\text{dist}_{im}$ (distance from $i$'s country to the host city, in $1{,}000$~km), $\text{jetlag}_{im}=|\text{tz}_i-\text{tz}_h|$ (absolute time zone difference in hours), and binary indicators for sharing the host's east-west hemisphere, north-south hemisphere, language, and religion. Table~\ref{tab:sumstats} in the Appendix reports descriptive statistics for all outcome and regressor variables in this panel.

\subsection{Econometric strategy}

Our workhorse specification is
\begin{equation}
y_{im}=\beta\,\text{prox}_{im}+\alpha_i+\gamma_j+\delta_t+\rho_s+\varepsilon_{im},
\label{eq:main}
\end{equation}
where $\text{prox}_{im}$ is one or more proximity measures, $\alpha_i$ is a focal-team fixed effect (absorbing the team's average strength), $\gamma_j$ an opponent fixed effect (absorbing the rival's strength), $\delta_t$ a year fixed effect, and $\rho_s$ a round fixed effect. Standard errors are clustered by tournament; results are robust to clustering by team and to heteroskedasticity-robust errors. The coefficient $\beta$ identifies the Regional Differential off \emph{within-team} variation in proximity across the venues a team visits, holding fixed both its own and its opponents' quality.

To test for heterogeneity we estimate team-specific gradients by interacting the proximity measure with a focal-team indicator,
\begin{equation}
y_{im}=\beta_k\,(\text{dist}_{im}\!\times\!\mathbf{1}[i=k])+\lambda\,\text{dist}_{im}+\theta_k\,\mathbf{1}[i=k]+\gamma_j+\delta_t+\rho_s+\varepsilon_{im},
\label{eq:hetero}
\end{equation}
where $\beta_k$ is team $k$'s own distance gradient, so that equation~\eqref{eq:hetero} is estimated for each team $k$. We report these estimates separately for every team with at least $100$ team-match observations, pooling genders and then splitting by gender.

\section{Empirical Results}

\subsection{Motivation: performance is not uniform across regions}

\begin{figure}[hbt!]
    \centering
    \includegraphics[width=0.72\linewidth]{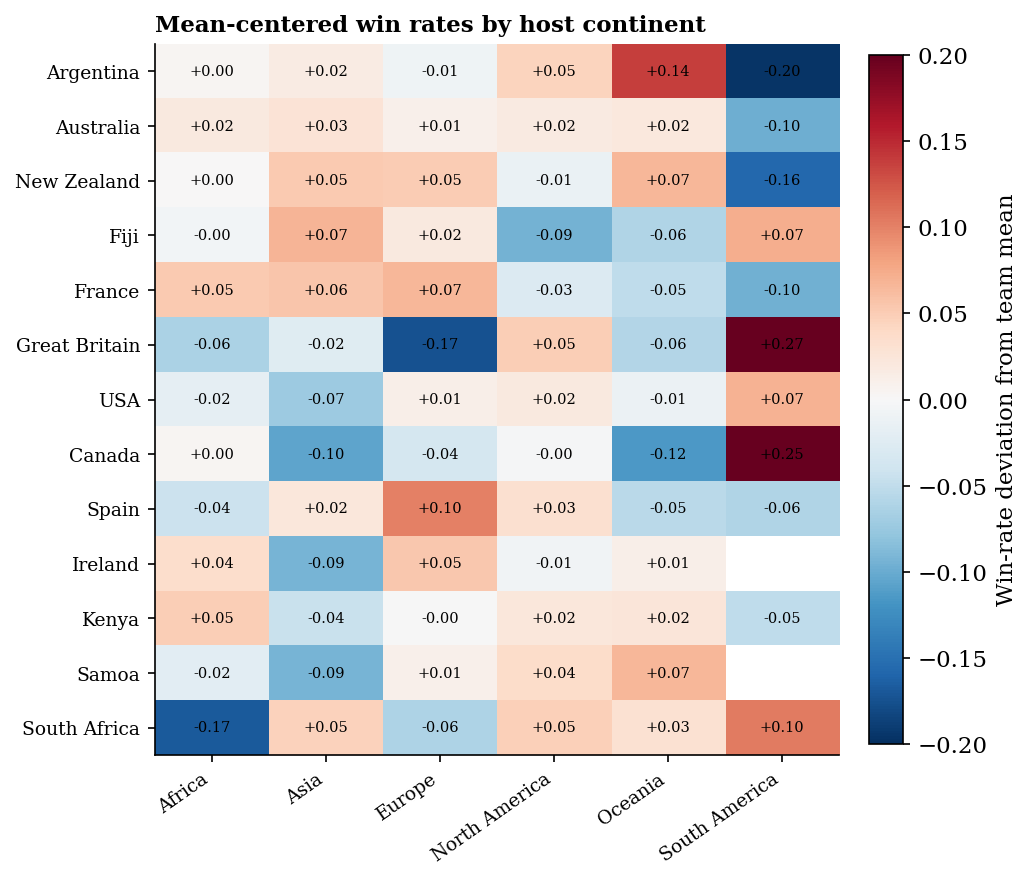}
    \caption{Mean-centered win rates by host continent. Cells show the deviation of a team's win rate in a given continent from its overall win rate. Blanks denote fewer than five matches.}
    \label{fig:heatmap}
\end{figure}

As a first, descriptive motivation, Figure~\ref{fig:heatmap} reports each team's win rate by host continent, expressed as a deviation from that team's overall win rate (so that rows sum to approximately zero). Clear patterns emerge: Argentina and New Zealand over-perform in Oceania and under-perform in South America. Great Britain and Canada display the opposite continental tilt. France leans towards Europe. These deviations are exactly the kind of regularity that fuels the folk belief in regional strongholds, and they raise the question that the rest of the paper addresses: are they a genuine venue effect, or an artifact of which teams meet which rivals where?

\subsection{The aggregate Regional Differential is a composition artifact}\label{sec:aggregate}

\begin{figure}[hbt!]
    \centering
    \includegraphics[width=0.8\linewidth]{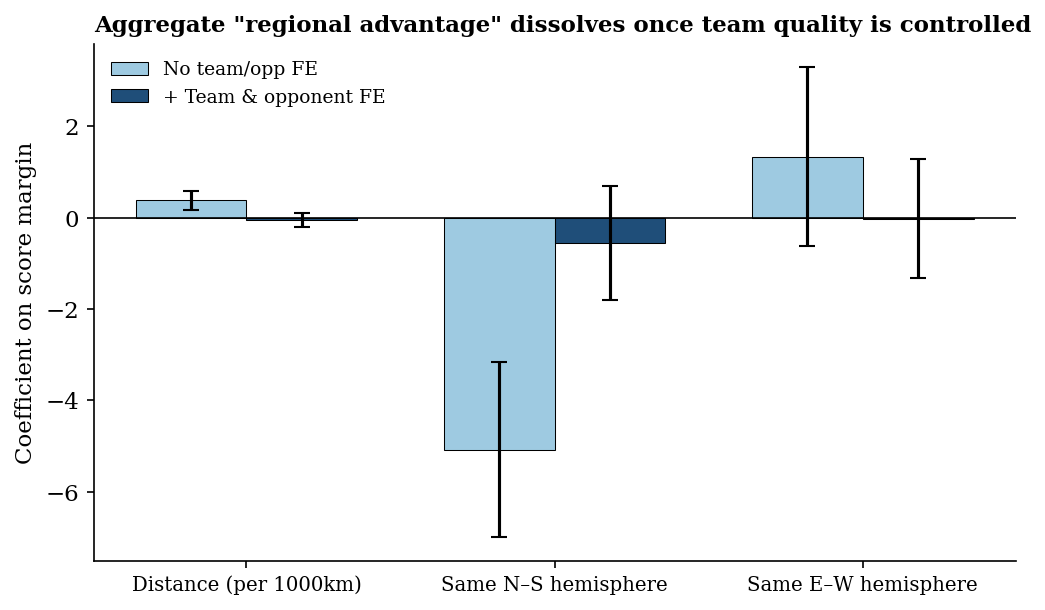}
    \caption{Selected proximity coefficients on score margin, with and without team and opponent fixed effects (95\% confidence intervals). The apparent aggregate effects vanish once quality is absorbed.}
    \label{fig:dissolution}
\end{figure}

Table~\ref{tab:pooled} estimates equation~\eqref{eq:main} with the full set of proximity measures. The left panel omits team and opponent fixed effects (retaining only year and round effects, as in conventional HA specifications). The right panel adds them. The contrast is stark. Without quality controls, sharing the host's north-south hemisphere is associated with a $4.6$-point \emph{lower} margin and sharing the east-west hemisphere with a $2.9$-point \emph{higher} margin (both highly significant). Jet lag itself carries a sizeable, highly significant positive coefficient, while shared language is only marginally significant and shared religion is not significant once the other proximity measures are held fixed; taken at face value, any of these estimates could be read as evidence of a ``regional advantage.'' Once team and opponent fixed effects enter, every one of these coefficients becomes statistically indistinguishable from zero (Figure~\ref{fig:dissolution}). The associations were driven by \emph{who} plays \emph{whom} and \emph{where}, not by a venue effect.

\begin{table}[hbt!]\centering\footnotesize
\caption{Regional-proximity effects on match outcomes (symmetric team--match panel)}\label{tab:pooled}
\begin{tabular}{l cccc cc}\toprule
 & \multicolumn{3}{c}{No team/opp.\ FE} & \multicolumn{3}{c}{Team \& opponent FE}\\
\cmidrule(lr){2-4}\cmidrule(lr){5-7}
 & Margin & Win & Win/tie & Margin & Win & Win/tie\\\midrule
Distance (1000 km) & -0.090 & -0.000 & -0.000 & -0.102 & -0.001 & -0.001\\
  & \scriptsize(0.121) & \scriptsize(0.003) & \scriptsize(0.003) & \scriptsize(0.125) & \scriptsize(0.003) & \scriptsize(0.003)\\
Jet lag (h) & 0.261$^{***}$ & 0.006$^{***}$ & 0.006$^{***}$ & -0.045 & -0.001 & -0.000\\
  & \scriptsize(0.080) & \scriptsize(0.002) & \scriptsize(0.002) & \scriptsize(0.072) & \scriptsize(0.002) & \scriptsize(0.002)\\
Same E-W hemisphere & 3.645$^{***}$ & 0.101$^{***}$ & 0.102$^{***}$ & -0.407 & 0.019 & 0.021\\
  & \scriptsize(0.934) & \scriptsize(0.019) & \scriptsize(0.018) & \scriptsize(0.772) & \scriptsize(0.016) & \scriptsize(0.015)\\
Same N-S hemisphere & -3.912$^{***}$ & -0.079$^{***}$ & -0.079$^{***}$ & -0.928 & -0.021 & -0.025\\
  & \scriptsize(0.934) & \scriptsize(0.019) & \scriptsize(0.018) & \scriptsize(0.844) & \scriptsize(0.018) & \scriptsize(0.018)\\
Same language & 1.650$^{**}$ & 0.012 & 0.017 & 0.095 & -0.004 & 0.001\\
  & \scriptsize(0.658) & \scriptsize(0.016) & \scriptsize(0.016) & \scriptsize(0.523) & \scriptsize(0.014) & \scriptsize(0.014)\\
Same religion & 2.657$^{***}$ & 0.039$^{**}$ & 0.043$^{**}$ & -0.463 & -0.019 & -0.015\\
  & \scriptsize(0.853) & \scriptsize(0.018) & \scriptsize(0.018) & \scriptsize(0.537) & \scriptsize(0.014) & \scriptsize(0.013)\\
\midrule Team \& opp.\ FE & No & No & No & Yes & Yes & Yes\\
Year \& stage FE & Yes & Yes & Yes & Yes & Yes & Yes\\
$N$ & 4821 & 4821 & 4821 & 4821 & 4821 & 4821\\
\bottomrule\end{tabular}
\caption*{\scriptsize SEs clustered by tournament. $^{*}p<0.1$, $^{**}p<0.05$, $^{***}p<0.01$.}
\end{table}

\subsection{Heterogeneity: who actually travels badly}

\begin{figure}[hbt!]
    \centering
    \includegraphics[width=0.72\linewidth]{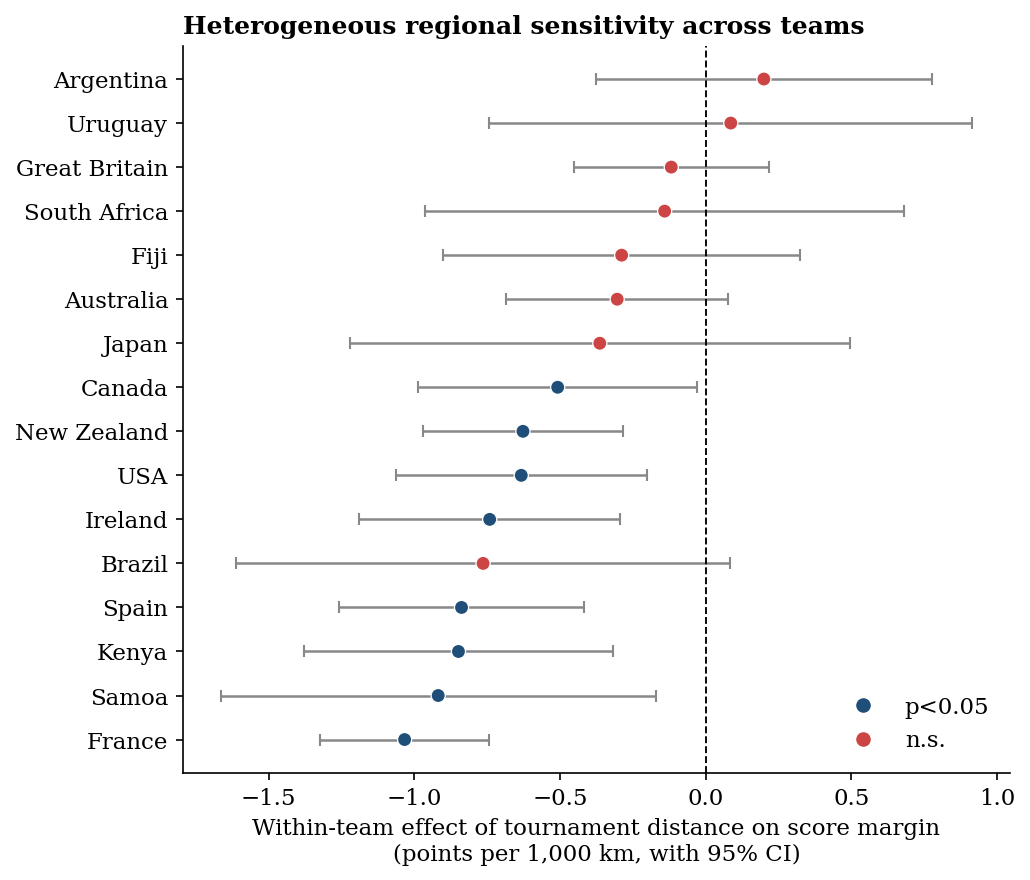}
    \caption{Within-team effect of host distance on score margin (points per $1{,}000$~km, $95\%$ CI), net of opponent, year, and round fixed effects. Dark markers: $p<0.05$.}
    \label{fig:coefplot}
\end{figure}

The aggregate null does not imply that proximity is irrelevant for \emph{every} team. Table~\ref{tab:teamgrad} reports team-specific distance gradients from equation~\eqref{eq:hetero} for all $16$ teams with at least $100$ team-matches, and Figure~\ref{fig:coefplot} plots the margin coefficients with confidence intervals.

\begin{table}[hbt!]\centering\footnotesize
\caption{Team-specific within-team distance gradients (effect of host distance, per 1000 km)}\label{tab:teamgrad}
\begin{tabular}{l c c c c}\toprule
 Team & Score margin & Win prob. & Win/tie prob. & Team-matches\\\midrule
France & -1.033$^{***}$ \scriptsize(0.148) & -0.018$^{***}$ \scriptsize(0.004) & -0.016$^{***}$ \scriptsize(0.004) & 389\\
Samoa & -0.918$^{**}$ \scriptsize(0.381) & -0.023$^{**}$ \scriptsize(0.012) & -0.025$^{**}$ \scriptsize(0.012) & 150\\
Kenya & -0.849$^{***}$ \scriptsize(0.271) & -0.012$^{*}$ \scriptsize(0.007) & -0.015$^{**}$ \scriptsize(0.007) & 194\\
Spain & -0.838$^{***}$ \scriptsize(0.215) & -0.016$^{***}$ \scriptsize(0.004) & -0.016$^{***}$ \scriptsize(0.004) & 353\\
Brazil & -0.764$^{*}$ \scriptsize(0.433) & -0.016$^{**}$ \scriptsize(0.008) & -0.016$^{**}$ \scriptsize(0.008) & 163\\
Ireland & -0.742$^{***}$ \scriptsize(0.228) & -0.011$^{*}$ \scriptsize(0.006) & -0.010$^{*}$ \scriptsize(0.006) & 382\\
USA & -0.633$^{***}$ \scriptsize(0.220) & -0.014$^{**}$ \scriptsize(0.005) & -0.015$^{***}$ \scriptsize(0.005) & 408\\
New Zealand & -0.627$^{***}$ \scriptsize(0.175) & -0.010$^{**}$ \scriptsize(0.005) & -0.010$^{**}$ \scriptsize(0.005) & 353\\
Canada & -0.508$^{**}$ \scriptsize(0.244) & -0.012$^{*}$ \scriptsize(0.006) & -0.012$^{*}$ \scriptsize(0.006) & 365\\
Japan & -0.364 \scriptsize(0.438) & 0.001 \scriptsize(0.009) & 0.000 \scriptsize(0.009) & 269\\
Australia & -0.304 \scriptsize(0.194) & -0.006 \scriptsize(0.004) & -0.005 \scriptsize(0.004) & 409\\
Fiji & -0.289 \scriptsize(0.312) & -0.001 \scriptsize(0.008) & -0.001 \scriptsize(0.008) & 371\\
South Africa & -0.141 \scriptsize(0.420) & -0.001 \scriptsize(0.010) & 0.000 \scriptsize(0.010) & 291\\
Great Britain & -0.118 \scriptsize(0.170) & -0.003 \scriptsize(0.005) & -0.004 \scriptsize(0.005) & 324\\
Uruguay & 0.086 \scriptsize(0.423) & -0.007 \scriptsize(0.010) & -0.007 \scriptsize(0.010) & 104\\
Argentina & 0.199 \scriptsize(0.294) & 0.004 \scriptsize(0.006) & 0.005 \scriptsize(0.006) & 235\\
\bottomrule\end{tabular}
\caption*{\scriptsize Each row: OLS of outcome on (team$\times$distance), team dummy, distance, opponent FE, year FE, stage FE; SEs clustered by tournament. Sorted by margin coefficient. $^{*}p<0.1$, $^{**}p<0.05$, $^{***}p<0.01$.}
\end{table}

\begin{figure}[hbt!]
    \centering
    \includegraphics[width=0.62\linewidth]{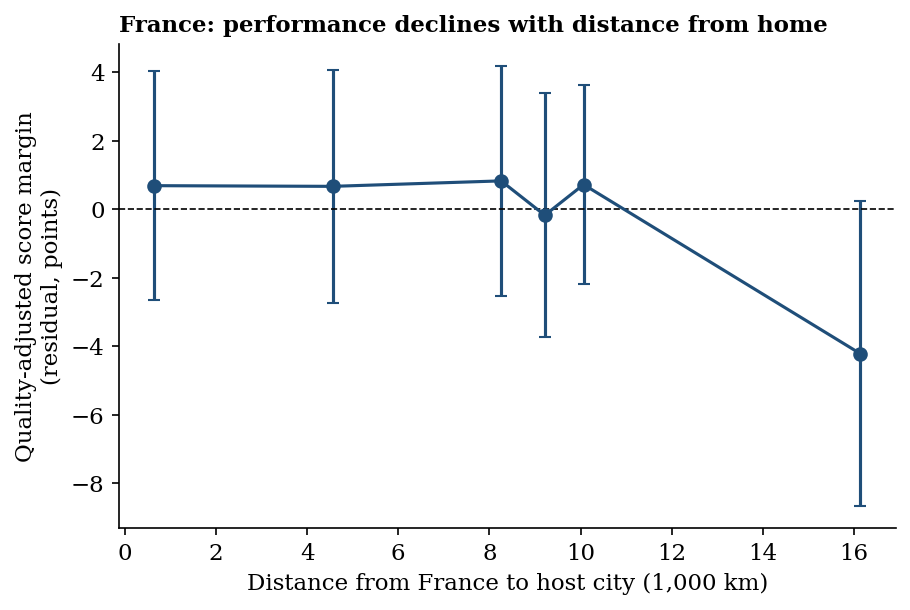}
    \caption{France: quality-adjusted score margin (residualized on opponent, year, and round) against distance to the host city, in six distance bins ($95\%$ CI).}
    \label{fig:france}
\end{figure}

Two regularities stand out. First, \emph{every} statistically significant gradient is negative: where distance matters, it always hurts, and no team performs reliably better the farther it travels. Second, the magnitude is highly heterogeneous. France shows the steepest decline, losing roughly one point of margin and $1.8$ percentage points of win probability per $1{,}000$~km. This is a substantial effect given that the farthest venues lie some $16{,}000$~km away (Figure~\ref{fig:france}). New Zealand, the United States, Canada, Spain, Ireland, Kenya, and Samoa display significant negative gradients of $0.5$--$0.9$ points; Brazil shows a comparable gradient ($-0.76$) that is significant at the $10\%$ level. By contrast, the established powers that are reputed to ``travel well'' (Fiji, South Africa, Australia, and Argentina), together with Great Britain, are statistically distance-neutral. The folk wisdom is thus half right: regional strongholds are real for some teams, but the teams most often credited with them (the southern-hemisphere greats) are precisely those for whom location does \emph{not} matter, whereas the effect is concentrated in a different, partly second-tier group, with the exception of New Zealand.

\subsection{Mechanisms: displacement versus jet lag}

Distance bundles together several channels. We separate pure displacement from east-west jet lag by including both $\text{dist}_{im}$ and $\text{jetlag}_{im}$, each interacted with the focal team, for the distance-sensitive sides. The channels differ across teams. For France the effect is pure distance: the displacement gradient remains $-1.05$ ($p<0.01$) while the jet-lag term is insignificant, consistent with France's worst results coming in far southern-hemisphere venues that are not especially time-shifted from Paris. New Zealand's gradient is likewise displacement-driven. Canada is the opposite case: conditioning on distance, its margin falls by $1.5$ points per hour of time difference ($p<0.01$), so Canada's apparent regional disadvantage is an \emph{east-west jet-lag} phenomenon, i.e.\ the channel emphasized by the circadian literature in other sports \citep{recht1995baseball,nutting2017timezones,leota2022eastward}, rather than one of raw distance. The United States lies in between, with neither distance nor jet lag reaching significance on its own. The Regional Differential, where present, therefore has team-specific physiology. That a displacement penalty should be physiologically real is consistent with direct evidence from rugby itself: long-haul transmeridian travel measurably degrades the sleep and recovery of professional rugby squads \citep{smithies2021around}, while in other elite team sports the eastward direction of travel has been singled out as especially harmful to performance and outcomes \citep{leota2022eastward}.

\subsection{Cultural proximity: a robust null}

If the Regional Differential reflected cultural familiarity rather than physical travel, we would expect teams to fare better at venues that resemble them culturally. They do not. Section~\ref{sec:aggregate} already showed that the binary indicators for sharing the host's language or religion, significant in raw form, vanish under team and opponent fixed effects (Table~\ref{tab:pooled}). Table~\ref{tab:culture} repeats the exercise with the finer, continuous measure: the absolute distance between team and host in each of the \citet{alesina2003fractionalization} fractionalization indices, a construction in the spirit of the cultural-distance literature \citep{spolaore2016ancestry}.

\begin{table}[hbt!]\centering\footnotesize
\caption{Cultural-distance effects on match outcomes (continuous fractionalization indices)}\label{tab:culture}
\begin{tabular}{l cc cc}\toprule
 & \multicolumn{2}{c}{No team/opp.\ FE} & \multicolumn{2}{c}{Team \& opp.\ FE}\\
\cmidrule(lr){2-3}\cmidrule(lr){4-5}
 & Margin & Win & Margin & Win\\\midrule
Ethnic distance & 4.49 & 0.02 & -3.64$^{**}$ & -0.08$^{**}$\\
 & \scriptsize(2.76) & \scriptsize(0.05) & \scriptsize(1.68) & \scriptsize(0.04)\\
Linguistic distance & -6.95$^{***}$ & -0.08$^{*}$ & 3.21$^{*}$ & 0.08$^{*}$\\
 & \scriptsize(2.60) & \scriptsize(0.05) & \scriptsize(1.75) & \scriptsize(0.04)\\
Religious distance & -2.89 & 0.01 & 0.87 & 0.03\\
 & \scriptsize(3.21) & \scriptsize(0.06) & \scriptsize(2.13) & \scriptsize(0.05)\\
\midrule Team \& opp.\ FE & No & No & Yes & Yes\\
Year \& stage FE & Yes & Yes & Yes & Yes\\
$N$ & 4990 & 4990 & 4990 & 4990\\
\bottomrule\end{tabular}
\caption*{\scriptsize Cultural distance is $|$team $-$ host$|$ in each Alesina et al.\ (2003) fractionalization index. SEs clustered by tournament. $^{*}p<0.1$, $^{**}p<0.05$, $^{***}p<0.01$.}
\end{table}

The result is a non-result. Without quality controls, linguistic distance carries a large, highly significant coefficient ($-6.9$, $p<0.01$). This is exactly the kind of estimate that could be paraded as evidence of a ``cultural'' home effect. Yet once team and opponent fixed effects enter, the coefficients are small, only marginally significant, and \emph{change sign and disagree with one another}: ethnic distance turns negative ($-3.6$, $p<0.05$) while linguistic distance turns positive ($+3.2$, $p<0.10$), with religious distance insignificant throughout. Because the three fractionalization dimensions are strongly collinear, this pattern is the signature of multicollinearity splitting an unstable, essentially zero effect across correlated regressors, not of a genuine cultural channel. Two further checks reinforce the null. First, host-level diversity cannot affect the score margin by construction in a symmetric panel (it is common to both teams and differences out) so any ``diverse-crowd'' mechanism is unidentifiable here. Second, interacting a team's \emph{own} ethnic diversity with distance, to test whether more cosmopolitan squads travel better, yields nothing ($\hat\beta=-0.09$, $p=0.78$). We therefore conclude that cultural proximity, measured as the similarity between the two countries' degrees of diversity, does not account for the Regional Differential. This sharpens the paper's central message: where regional performance gaps exist, they are about physical displacement, not about homogeneous versus heterogeneous attendance. We report the null explicitly because ruling out the culturalist explanation is itself informative for a literature that often invokes it.

\subsection{Men and women}

Splitting equation~\eqref{eq:hetero} by gender, the qualitative pattern is preserved. France's distance gradient is large and significant for both men ($-0.96$, $p<0.01$) and women ($-1.06$, $p<0.01$). New Zealand and Canada are significant in both. For the United States and Australia the gradient is concentrated on the women's side ($-0.92$ and $-0.42$ respectively, both $p<0.05$), while their men's coefficients are insignificant. Argentina fields no women's team in the sample. Overall, regional sensitivity is not a male or a female phenomenon. It is a team phenomenon that, for a few federations, manifests more strongly in one gender than in the other.

\subsection{Robustness}

The team-specific gradients are not an artifact of the inference choice. Table~\ref{tab:robust} re-estimates the nine significant gradients of Table~\ref{tab:teamgrad} under three error structures---standard errors clustered by tournament (the baseline), clustered by team, and heteroskedasticity-robust (HC1)---and additionally drops the year fixed effects. The point estimates are unchanged by construction, and the statistical conclusions are stable: every gradient remains significant at conventional levels across all inference methods, with the strongest cases (France, New Zealand, Spain, Kenya) significant at $p<0.01$ throughout. The only sensitivities are Canada, whose gradient weakens to $p=0.082$ when year effects are removed but tightens (to $p=0.014$ and $p=0.007$) under team-clustering and HC1, respectively, and Brazil, which is significant only at the $10\%$ level under tournament-clustering ($p=0.082$) and when year effects are dropped ($p=0.085$), but tightens to $p=0.001$ under team-clustering and $p=0.016$ under HC1. The aggregate null of Section~\ref{sec:aggregate} is likewise robust: the pooled distance coefficient with team and opponent fixed effects is $-0.05$ whether errors are clustered by tournament ($p=0.50$) or by team ($p=0.40$). We thus report the within-team distance gradient as a stable feature of the data rather than a fragile finding.

\begin{table}[hbt!]\centering\footnotesize
\caption{Robustness of the team-specific distance gradient ($p$-values under alternative inference)}\label{tab:robust}
\begin{tabular}{l c cccc}\toprule
Team & $\hat\beta$ (margin) & Cluster: tourn. & Cluster: team & HC1 robust & Drop year FE\\\midrule
France & -1.033 & 0.000$^{***}$ & 0.000$^{***}$ & 0.000$^{***}$ & 0.000$^{***}$\\
New Zealand & -0.627 & 0.000$^{***}$ & 0.000$^{***}$ & 0.000$^{***}$ & 0.000$^{***}$\\
USA & -0.633 & 0.004$^{***}$ & 0.003$^{***}$ & 0.000$^{***}$ & 0.004$^{***}$\\
Canada & -0.508 & 0.037$^{**}$ & 0.014$^{**}$ & 0.007$^{***}$ & 0.082$^{*}$\\
Spain & -0.838 & 0.000$^{***}$ & 0.000$^{***}$ & 0.000$^{***}$ & 0.000$^{***}$\\
Ireland & -0.742 & 0.001$^{***}$ & 0.000$^{***}$ & 0.000$^{***}$ & 0.002$^{***}$\\
Kenya & -0.849 & 0.002$^{***}$ & 0.000$^{***}$ & 0.001$^{***}$ & 0.000$^{***}$\\
Samoa & -0.918 & 0.016$^{**}$ & 0.000$^{***}$ & 0.003$^{***}$ & 0.008$^{***}$\\
\bottomrule\end{tabular}
\caption*{\scriptsize Each cell is the $p$-value for the focal team's distance gradient (per $1{,}000$ km) from the specification of Table~\ref{tab:teamgrad}, under the indicated inference method or, in the last column, dropping year fixed effects. $\hat\beta$ is the point estimate (identical across inference methods). $^{*}p<0.1$, $^{**}p<0.05$, $^{***}p<0.01$.}
\end{table}

\section{Discussion}

Four lessons follow. The first is methodological and, we believe, of general interest to the sports economics literature on neutral-venue competition: in settings without formal home teams, ``locality'' must be treated as continuous \emph{and} estimated within a design that absorbs the quality of both contestants. The cross-sectional correlations that underpin descriptive accounts of regional strongholds are driven by composition (strong teams of a given region meet weaker rivals at certain venues) and evaporate under team and opponent fixed effects. Studies that report neutral-venue ``home'' effects without such controls risk mistaking the schedule for the stadium.

Second, the heterogeneity is the substantive finding. A genuine, robust distance penalty exists for a recognizable group of teams, and it is economically meaningful: for France it is comparable in size to the round-of-competition effect. That the penalty is absent for Fiji, South Africa, Australia, and Argentina, while present for France, New Zealand, the United States, Canada, Spain, Ireland, Kenya, Samoa, and Brazil, suggests an explanation rooted in squad depth, professionalization, and acclimatization resources rather than in geography \emph{per se} (with the possible exception of New Zealand): the sides able to neutralize travel are, broadly, those with the deepest sevens teams and the longest history of intercontinental campaigning.

Third, the channel is not monolithic. The contrast between France (pure displacement) and Canada (east-west jet lag) indicates that ``travel'' enters performance through at least two distinct physiological routes, a distinction the binary HA framework cannot represent and that future work could pursue with richer biometric or itinerary data.

The fourth lesson is what does \emph{not} matter. We motivated a cultural channel from the economics-of-culture tradition (the fractionalization indices of \citealt{alesina2003fractionalization} and the cultural-distance arguments of \citealt{spolaore2013roots,spolaore2016ancestry}) under the natural hypothesis that teams enjoy playing in similarly diverse venues. The data reject it. The continuous distance in any fractionalization index does not have a stable association with results once quality is controlled, and the only sizeable raw coefficient flips sign and disagrees with its companions under fixed effects (Table~\ref{tab:culture}). The home advantage premium, in other words, does not appear to travel through cultural proximity in this setting. Where regional performance gaps survive, they are about physical displacement, not cultural familiarity.

Several limitations temper these conclusions. Flight-time and connection data are incomplete and were therefore not used in the main specifications. While the panel design removes the most obvious confounders, residual selection (e.g.\ rotation policies that rest stars on long trips) cannot be fully excluded. We view the team-specific gradients as well-identified descriptions of how results co-move with travel within each team's schedule, instead of estimates of a structural fatigue parameter.

\section{Conclusion}

In this work we asked whether a home advantage can exist when nobody is home. Using the international rugby sevens competitions during 2016--2025, we showed that the much-discussed Regional Differential is, in the aggregate, a statistical illusion produced by team composition and scheduling: it disappears once team and opponent quality are absorbed. Yet beneath that null lies a robust and economically meaningful heterogeneity. A specific set of teams, most sharply France, but also New Zealand, the United States, Canada, Spain, Ireland, Kenya, Samoa, and Brazil, performs worse the farther a tournament is staged from home, at $0.5$--$1.0$ points of margin per $1{,}000$~km, while the remaining traditional powers are immune. For some teams the penalty is one of distance, for others of time zone displacement. The Regional Differential in sevens is therefore best understood not as a property of the sport but as a property of particular teams: a few federations carry their geography with them, and most do not.

\appendix
\section{Appendix}
\subsection{Descriptive statistics}

Table~\ref{tab:sumstats} reports descriptive statistics for the outcome and regressor variables used in Tables~\ref{tab:pooled}--\ref{tab:host}, computed on the symmetric team-match panel described in Section~\ref{sec:panel}.

\begin{table}[hbt!]\centering\footnotesize
\caption{Descriptive statistics, symmetric team-match panel}\label{tab:sumstats}
\begin{threeparttable}
\begin{tabular}{l c c c c c}\toprule
Variable & $N$ & Mean & SD & Min & Max \\\midrule
\multicolumn{6}{l}{\textit{Outcomes}} \\
Score margin & 5344 & 0.000 & 20.792 & -70.000 & 70.000 \\
Win (=1) & 5344 & 0.495 & 0.500 & 0 & 1 \\
Win or tie (=1) & 5344 & 0.505 & 0.500 & 0 & 1 \\
\multicolumn{6}{l}{\textit{Geographic and temporal proximity}} \\
Distance (1{,}000 km) & 5344 & 9.207 & 4.687 & 0.000 & 19.855 \\
Jet lag (h) & 5344 & 6.926 & 5.105 & 0.000 & 21.000 \\
Same E-W hemisphere (=1) & 5344 & 0.514 & 0.500 & 0 & 1 \\
Same N-S hemisphere (=1) & 5344 & 0.547 & 0.498 & 0 & 1 \\
\multicolumn{6}{l}{\textit{Cultural proximity}} \\
Same language (=1) & 5344 & 0.371 & 0.483 & 0 & 1 \\
Same religion (=1) & 5344 & 0.391 & 0.488 & 0 & 1 \\
Ethnic distance & 4990 & 0.289 & 0.208 & 0.000 & 0.847 \\
Linguistic distance & 4990 & 0.301 & 0.219 & 0.000 & 0.868 \\
Religious distance & 4990 & 0.216 & 0.175 & 0.000 & 0.716 \\
\bottomrule\end{tabular}
\begin{tablenotes}\footnotesize
\item[] \textit{Notes:} Symmetric team-match panel built from the $2{,}672$ matches described in Section~\ref{sec:data} ($N=5{,}344$ team-match rows). Score margin is the focal team's score minus the opponent's. Distance is the focal team's distance to the host city. Jet lag is the absolute difference in standard-time UTC offset between the focal team and the host, complete for all rows (see Table~\ref{tab:pooled}). Same E-W/N-S hemisphere, same language, and same religion are binary indicators for the focal team sharing that attribute with the host. Ethnic, linguistic, and religious distance are $|$focal team $-$ host$|$ in the corresponding \citet{alesina2003fractionalization} fractionalization index (range $0$--$1$), available for a subset of matches with non-missing fractionalization data (see Table~\ref{tab:culture}).
\end{tablenotes}
\end{threeparttable}
\end{table}

\subsection{Hemisphere specifications}

For comparability with the home advantage literature's hemisphere-based treatments of locality, Tables~\ref{tab:appdif1}--\ref{tab:appdif2} report match-level (one row per match, not the symmetric team-match panel of the main text) score-difference regressions on east-west and north-south hemisphere indicators, computed separately for Team~1 and Team~2 against the host. Team~1 and Team~2 here follow the raw match listing described in Section~\ref{sec:data} (not the focal-team convention of the rest of the paper). Because that listing is not symmetric, these regressions are offered only as a bridge to the traditional literature. The coefficients on Team~1 and Team~2 are not directly comparable to each other. For consistency with the rest of the paper, indicators are coded as \emph{Same} hemisphere (as in Table~\ref{tab:pooled}), not \emph{different} hemisphere. Table~\ref{tab:appdif1} considers the east-west hemisphere alone; Table~\ref{tab:appdif1b} enters the east-west and north-south indicators jointly. Consistent with Section~\ref{sec:aggregate}, the apparent hemisphere effects in these specifications, which omit team and opponent fixed effects, are substantially attenuated or eliminated in the panel of Table~\ref{tab:pooled}.

\begin{table}[hbt!]\centering\footnotesize
\caption{East-west hemisphere differentials on the score difference, full sample}\label{tab:appdif1}
\begin{threeparttable}
\begin{tabular}{l cc}\toprule
 & (1) No controls & (2) Full controls \\\midrule
Same E-W hemisphere (Team 1) & 2.86$^{***}$ & 2.54$^{***}$ \\
Same E-W hemisphere (Team 2) & 1.07 & 1.01 \\
Same language (Team 1) & -- & -0.43 \\
Same language (Team 2) & -- & 3.50$^{**}$ \\
Same religion (Team 1) & -- & 0.27 \\
Same religion (Team 2) & -- & 1.14 \\
\midrule
Year \& round FE & No & Yes \\
$N$ & 2672 & 2672 \\
\bottomrule\end{tabular}
\begin{tablenotes}\footnotesize
\item[] \textit{Notes:} Match-level OLS (one row per match, $N=2{,}672$) of the score difference (Team~1 $-$ Team~2) on indicators for whether each team shares the host's east-west hemisphere, heteroskedasticity-robust standard errors. Column (1) has no additional controls. Column (2) adds an indicator for each team sharing the host's dominant language, an indicator for each team sharing the host's dominant religion, and year and round fixed effects (coefficients on individual year/round dummies estimated but not tabulated). $^{*}p<0.1$, $^{**}p<0.05$, $^{***}p<0.01$.
\end{tablenotes}
\end{threeparttable}
\end{table}

\begin{table}[hbt!]\centering\footnotesize
\caption{East-west and north-south hemisphere differentials, joint specification, full sample}\label{tab:appdif1b}
\begin{threeparttable}
\begin{tabular}{l cc}\toprule
 & (1) No controls & (2) Full controls \\\midrule
Same N-S hemisphere (Team 1) & -2.73$^{***}$ & -2.55$^{***}$ \\
Same N-S hemisphere (Team 2) & 5.00$^{***}$ & 4.28$^{***}$ \\
Same E-W hemisphere (Team 1) & 2.93$^{***}$ & 2.62$^{***}$ \\
Same E-W hemisphere (Team 2) & 0.76 & 0.75 \\
Same language (Team 1) & -- & -0.06 \\
Same language (Team 2) & -- & 3.51$^{**}$ \\
Same religion (Team 1) & -- & 0.32 \\
Same religion (Team 2) & -- & 0.72 \\
\midrule
Year \& round FE & No & Yes \\
$N$ & 2672 & 2672 \\
\bottomrule\end{tabular}
\begin{tablenotes}\footnotesize
\item[] \textit{Notes:} Match-level OLS (one row per match, $N=2{,}672$) of the score difference (Team~1 $-$ Team~2) on the east-west and north-south hemisphere indicators entered jointly (unlike Table~\ref{tab:appdif1}, which considers east-west alone), heteroskedasticity-robust standard errors. Column (1) has no additional controls; column (2) adds the same language, religion, year, and round controls as Table~\ref{tab:appdif1}. $^{*}p<0.1$, $^{**}p<0.05$, $^{***}p<0.01$.
\end{tablenotes}
\end{threeparttable}
\end{table}

\begin{table}[hbt!]\centering\footnotesize
\caption{Hemisphere differentials on the score difference, by gender (no controls)}\label{tab:appdif2}
\begin{threeparttable}
\begin{tabular}{l cc}\toprule
 & Men & Women \\\midrule
Same N-S hemisphere (Team 1) & -2.84$^{***}$ & -3.08$^{**}$ \\
Same N-S hemisphere (Team 2) & 5.60$^{***}$ & 3.91$^{***}$ \\
Same E-W hemisphere (Team 1) & 0.86 & 5.41$^{***}$ \\
Same E-W hemisphere (Team 2) & 1.71$^{*}$ & 0.33 \\
\midrule
$N$ & 1600 & 1072 \\
\bottomrule\end{tabular}
\begin{tablenotes}\footnotesize
\item[] \textit{Notes:} Match-level OLS of the score difference (Team~1 $-$ Team~2) on the hemisphere indicators, estimated separately by gender, with no additional controls (heteroskedasticity-robust standard errors); the north-south and east-west indicators come from separate regressions. $^{*}p<0.1$, $^{**}p<0.05$, $^{***}p<0.01$.
\end{tablenotes}
\end{threeparttable}
\end{table}

\subsection{Per-host-country gradients}

Table~\ref{tab:host} reorganizes the distance evidence by venue rather than by team. For each host country we take the quality-adjusted residual from a global model with team, opponent, year, and round fixed effects, and regress it on the focal team's distance to that host among matches played there. A negative gradient means that, relative to their own baseline, teams underperform the farther they have traveled to reach that particular venue. The penalty is concentrated at a handful of hosts: it is strongest and statistically detectable in the United States, the United Arab Emirates, and Spain, essentially flat at most venues, and reverses in South Africa, where the long-haul visitors who do make the trip tend to over-perform. We stress that within a single host distance varies only across teams, so these gradients blend proximity with any host-specific selection of which teams attend; they are descriptive companions to the well-identified within-team estimates of Table~\ref{tab:teamgrad}.

\begin{table}[hbt!]\centering\footnotesize
\caption{Per-host-country distance gradients on quality-adjusted performance}\label{tab:host}
\begin{tabular}{l c c c}\toprule
Host country & Dist.\ gradient (margin/1000 km) & Mean team distance (1000 km) & Team-matches\\\midrule
United States & -0.403$^{*}$ \scriptsize(0.206) & 9.4 & 606\\
UAE & -0.331$^{**}$ \scriptsize(0.162) & 9.1 & 650\\
Brazil & -0.254 \scriptsize(0.367) & 9.4 & 136\\
Spain & -0.241$^{*}$ \scriptsize(0.143) & 6.1 & 434\\
New Zealand & -0.212 \scriptsize(0.169) & 10.8 & 156\\
Australia & -0.121 \scriptsize(0.148) & 11.2 & 410\\
Singapore & -0.002 \scriptsize(0.250) & 10.5 & 404\\
France & 0.022 \scriptsize(0.112) & 8.2 & 452\\
Hong Kong & 0.076 \scriptsize(0.171) & 10.1 & 502\\
Japan & 0.086 \scriptsize(0.299) & 9.0 & 134\\
United Kingdom & 0.122 \scriptsize(0.128) & 8.4 & 172\\
Canada & 0.236 \scriptsize(0.162) & 8.5 & 748\\
South Africa & 0.451$^{**}$ \scriptsize(0.222) & 9.9 & 540\\
\bottomrule\end{tabular}
\caption*{\scriptsize Dependent variable: quality-adjusted residual from a global model with team, opponent, year and round fixed effects. Each row regresses that residual on the focal team's distance to the host, among matches played in that host country (heteroskedasticity-robust SEs). A negative gradient means teams underperform their own baseline the farther they travel to that venue. Within a host, distance varies only across teams, so the gradient is descriptive. $^{*}p<0.1$, $^{**}p<0.05$, $^{***}p<0.01$.}
\end{table}

\noindent The cleaned data are available in the project repository: \url{https://bit.ly/3P4PFW4}.

\bibliography{ref}
\end{document}